\documentclass[%
 reprint,
 amsmath,amssymb,
 aps,
prb,
]{revtex4-2}

\usepackage{graphicx}
\usepackage{dcolumn}
\usepackage{bm}

\usepackage{tabularx} 
\usepackage{enumerate}
\usepackage{txfonts}
\usepackage[font=small,labelfont=bf]{caption}
\usepackage{pgfplots}
\usetikzlibrary{calc}
\usepackage{extarrows}
\usepackage{booktabs,subcaption,amsfonts,dcolumn}
\definecolor{refkey}{rgb}{0.9451,0.2706,0.4941}
\definecolor{labelkey}{rgb}{0.9451,0.2706,0.4941}
\DeclareFontFamily{U}{rcjhbltx}{}
\DeclareFontShape{U}{rcjhbltx}{m}{n}{<->rcjhbltx}{}
\DeclareSymbolFont{hebrewletters}{U}{rcjhbltx}{m}{n}
\DeclareMathSymbol{\tet}{\mathord}{hebrewletters}{84}
\DeclareMathSymbol{\pey}{\mathord}{hebrewletters}{112}

\def\z2{$\mathbb{Z}_2$}

\definecolor{darkgray}{rgb}{0.33, 0.33, 0.33}
\usepackage{braket}
\newcommand{\kket}[1]{\left|\left.#1\right\rangle\!\right\rangle}
\newcommand{\bket}[1]{{\Ket{\widehat{#1}}}}
\newcommand{\bbra}[1]{\left\langle\!\left\langle#1\right|\right.}
\newcommand{\Tr}{{\mathrm{Tr}}}
\newcommand{\fig}[1]{\mathop{\vcenter{\hbox{{\includegraphics[width=1.25cm]{#1.png}}}}}\limits}
\newcommand{\textfig}[1]{\mathop{\vcenter{\hbox{{\includegraphics[width=0.7cm]{#1.png}}}}}\limits}
\newcommand{\abs}[1]{\left|#1\right|}

\begin{document}


\title{Fermionization of conformal boundary states}

\author{Hiromi Ebisu}
\email{hiromi.ebisu@weizmann.ac.il}
\altaffiliation{Department of Condensed Matter Physics}
\author{Masataka Watanabe}%
\email{masataka.watanabe@weizmann.ac.il}
\altaffiliation{Department of Particle Physics and Astrophysics}
\affiliation{
Weizmann Institute of Science, Rehovot, 7610001, Israel
}






\begin{abstract}
    We construct the complete set of boundary states of two-dimensional fermionic CFTs using that of the bosonic counterpart.
    We see that there are two groups of boundary conditions, which contributes to the open-string partition function by characters with integer coefficients, or with $\sqrt{2}$ times integer coefficients.
    We argue that, using the argument of [JHEP 09 (2020) 018], this $\sqrt{2}$ indicates a single unpaired Majorana zero mode, and that these two groups of boundary conditions are mutually incompatible.
    We end the paper by mentioning a possible interpretation of the result in terms of the entanglement entropy.
\end{abstract}

\maketitle




\newpage

\section{Introduction}

Bosonization is a procedure to get an equivalent bosonic theory out of a fermionic one \cite{PhysRevD.11.2088,Gogolin}, or fermionization, vice versa.
They find many interesting applications in condensed matter as well as high-energy physics, serving for better understanding of systems with strongly interacting systems including fermions.
One primitive such example is the Luttinger liquid \cite{Luttinger}, where one finds a description of an interacting theory of one-dimensional fermions as a theory of free bosons.
They have also been attracting attention recently in the context of fermionic symmetry protected topological (SPT) phases \cite{1201.2648,PhysRevB.83.035107,PhysRevB.83.075103,1505.05856,1605.01640,1908.04805,1911.11780}.

Bosonization/fermionization are understood the simplest in $1+1$ dimensions \cite{1902.05550}.
In two-dimensions, the bosonization is simply a sum of spin structures, which will yield a bosonic theory with non-anomalous \z2 symmetry.
Conversely, the fermionization is possible whenever the bosonic theory has non-anomalous \z2 symmetry, on which one tensors the Arf invariant (\it i.e., \rm the Kitaev chain~\cite{kitaev2001unpaired}), and then gauges the common \z2 subgroup \cite{1902.05550,2002.12283}.\footnote{
For more information and for a complete list of minimal models which can be fermionised, the reader is referred to \cite{2002.12283,2003.04278}.}
Of particular importance among such theories are the Ising ($c=1/2$) and the tricritical Ising ($c=7/10$) models, respectively mapping to a theory of free Majorana fermion and the smallest $\mathcal{N}=1$ minimal model.

One can also bosonize or fermionize a theory on a manifold with boundaries.
In this case, one also needs to map boundary conditions under the duality --
This has been studied in two-dimensions in the context of open string worldsheet ending on D-branes, and as well as in higher-dimensions in the context of particle-vortex duality of Chern-Simons-matter theories \cite{hep-th/0201173,1712.07654,1712.02801,1803.08507}.

Incidentally, note that the matching of boundary conditions across duality will have direct physical consequence, for example on the computation of the R\'enyi entanglement entropy.
It was pointed out that the computation of such an object crucially depends on the boundary condition between subregions (of size $L$) and its compliment and that the effect becomes visible when the number of replica index $n$ becomes of the same order as $\log\left(\frac{L}{\epsilon_{\rm UV}}\right)$ in \cite{1002.4353,1406.4167,2101.03320}.
It was also pointed out that one needs to be careful about the change in boundary conditions, if one wishes to match the reduced density matrix across duality \cite{1605.09396,1808.05939}.

In spite of being equivalent, boundary conditions of fermionic theories are more interesting than that of the bosonic theories, because of the notion of mutual compatibility of boundary conditions once we put the theory on the cylinder.
This can already be seen for a simple theory like a free Majorana fermion -- 
There are two boundary conditions $V:\chi_L=\chi_R$ and $A:\chi_L=-\chi_R$, and if one puts $V$ and $A$ boundary conditions on each end of the cylinder, this becomes inconsistent because of the presence of a single Majorana zero mode (See for example \cite{10.1143/PTPS.177.42} and references therein.).
For a more general case of $N$ Dirac fermions, \cite{1912.01602,2005.11314,2006.07369} classified the boundary conditions into two equivalence classes of mutually inconsistent ones using open-closed modular bootstrap.\footnote{For more information about the open-closed modular bootstrap programme, the reader is referred to \cite{1206.5395,1305.2122}.}


Having said that, the main question we are going to ask in this paper is the following: Can we construct the complete set of boundary conditions of two-dimensional fermionic conformal field theories (CFTs), when one is given the complete set of boundary conditions of two-dimensional bosonic CFTs?
In two-dimensional CFTs, one can use the open-closed duality and quantify the question in terms of boundary states \cite{hep-th/0411189}.
In this language, we are going to find all the elementary boundary states of fermionic CFTs, with a slight caveat that the definition of elementary here be generalised so that the open-string loop consists not only of characters with integer coefficients, but also $\sqrt{2}$ times integers, indicating that we allow for a real fermion on the boundary.
We are going to call this ``generalised elementary'' in this paper.

Fermionization is similar in spirit to orbifolding in that they are both a sum of \z2 defect configurations with particular coefficients.
Thus, we will also discuss the \z2 orbifolding of boundary states along with fermionization, although the former has been studied many times before in the context of string worldsheet {\cite{hep-th/0201173,hep-th/0108126,hep-th/9708141,hep-th/0011060}}.
This is not just being pedagogical, and we will see that comparing fermionization against orbifolding also serves as a small consistency check. 
When we present the final result, we are going to see in a clear way how boundary states from different sectors are shuffled among themselves when we orbifold, bosonize, or fermionize the theory.



The rest of the paper is organised as follows.
In Section~\ref{ising}, we briefly review boundary conformal field theories (BCFTs).
We will also study the bosonization/fermmionization of boundary states for the Ising model (or equivalently for the theory of free Majorana fermion).
In Section~\ref{general}, we determine the complete set of elementary boundary states for fermionic and orbifold CFTs in terms of the original boundary states.
We conclude in Section~\ref{sec:discussion}, mentioning possible interesting future directions. 

\section{Preliminary}\label{ising}
As a preliminary, we review basic notions of BCFT and
how bosonization/fermionization in BCFT works in a simple example, $c=1/2$ CFT, which might be a hint for consideration in later sections. This procedure was discussed in \cite{Bachas:2012bj} in the context of the GSO projection of D-branes.\par
\subsection{Elementary boundary states}
Conformal boundary condition reads 
\begin{equation}
    T(z)=\overline{T}(\overline{z})\;\; \text{at}\;z=\overline{z},
\end{equation}
where $T(z)$ (resp. $\overline{T}(\overline{z})$) is the stress-energy tensor in holomorphic (resp. anti-holomorphic) sector. This condition is mapped to
\begin{equation}
    (L_n-\overline{L}_{-n})\kket{B}=0,\;\;n\in \mathbb{Z}.\label{glue}
\end{equation}
Here, $L_n$ (resp. $\overline{L}_{-n}$) is the Laurent mode of the stress-energy tensor in the holomorphic (resp. anti-holomorphic) sector, and $\kket{B}$ is a boundary state. The solutions of (\ref{glue}) is known, which are the so-called Ishibashi states~\cite{ishibashi}. 
Generally, the Ishibashi states are not physical; they do not satisfy a consistency condition, which is called the Cardy condition [see (\ref{cardy condition}) below]. To have the physical boundary states, we introduce a boundary states which is a linear combination of the Ishibashi states denoted by $\kket{i}$ with $i$ corresponding to a Virasoro representation of a highest weight $\phi_i$,
\begin{equation}
    \bket{a}=\sum_i B^a_i\kket{i}\label{car}.
\end{equation}
Consider a cylinder partition function (closed string partition function) between two boundary states, $\bket{a}$ and $\bket{b}$ 
\begin{equation}
   Z^{\text{closed}}=\Braket{\widehat{a}|q^{\frac{1}{2}\left(L_0+\bar{L}_0-\frac{c}{12}\right)}|\widehat{b}}=\sum_i\overline{B^a_i}B^b_i\chi_i(q).
\end{equation}
Here, $q$ is written as $q\equiv e^{-\frac{4\pi L}{\beta}}$, using the length $L$ and the inverse temperature $\beta$ of the cylinder, and $\chi_i(q)$ is the Virasoro character. We also have used a relation 
$ \bbra{i}{q}^{\frac{1}{2}(L_0+\overline{L}_0-1/24)}\kket{j}=\delta_{i,j}\chi_i(q)$. An open string partition function is obtained by $S$-transforming the closed string partition function, which should have the form 
\begin{equation}
    Z^{\text{open}}_{ab}=\sum_in_{ab}^i\chi_i(\tilde{q})\label{open}
\end{equation}
with $n_{ab}^i$ being non-negative integer and $\tilde{q}\equiv e^{-\frac{\pi\beta}{L}}$ the $S$-transformation of $q$. Since two partition functions are related via $S$-transformation, one obtains
\begin{equation}
   \sum_i\overline{B^a_i}B^b_iS_{ij}\chi_j(q) =\sum_in_{ab}^i\chi_i(\tilde{q})\label{cardy condition}.
\end{equation}
The condition~(\ref{cardy condition}) is called the Cardy condition
and the boundary state~(\ref{car}) satisfying this condition is called the Cardy state\footnote{Throughout this paper, the double-ket represents the Ishibashi state whereas the ket with hat denotes the Cardy state. }~\cite{cardy1989boundary}. In the case of a charge conjugate CFT, 
if we set the coefficients in (\ref{car}) to be $B^a_i=\frac{S_{ai}}{\sqrt{S_{0i}}}$ with $S_{ab}$ being modular S-matrix, one can verify the boundary state satisfies the Cardy condition by use of the Verlinde formula.

Let us introduce a notion of elementary boundary states~\cite{hep-th/0108238}, which is crucial in our paper. 
A complete set of elementary boundary states is collection of boundary states, $\left\{\bket{a_n}\right\}$, among which we have $n^{0}_{a_i a_j}=\delta_{ij}$.
In other words, when an elementary boundary condition is placed on both ends of a cylinder, the open-string loop contains one vacuum character with unit coefficient, whereas if two different elementary boundary conditions are placed on two ends of a cylinder, the loop does not contain the vacuum character.


\subsection{Ising BCFT}
In the case of the Ising CFT, there are three Ishibashi states denoted by $\kket{0}$, $\kket{1/2}$, and $\kket{1/16}$, corresponding to three primary fields with conformal weight $0$, $1/2$, and $1/16$. The Cardy states 
are given by
\begin{eqnarray}
    \bket{+}_A&=&\frac{1}{\sqrt{2}}\kket{0}+\frac{1}{\sqrt{2}}\kket{1/2}+\frac{1}{\sqrt[4]{2}}\kket{1/16}\notag\\
        \bket{-}_A&=&\frac{1}{\sqrt{2}}\kket{0}+\frac{1}{\sqrt{2}}\kket{1/2}-\frac{1}{\sqrt[4]{2}}\kket{1/16}\notag\\
    \bket{f}_A&=&\kket{0}-\kket{1/2}.\label{boson ising}
\end{eqnarray}
Also, these states are elementary.
Physically, the first, second, and third boundary states correspond to a fixed boundary state with spin up, the one with spin down, and a free boundary state of the transverse Ising chain, respectively. If we act $\mathbb{Z}_2$ symmetry operator, physically corresponding to the spin flip, the first two boundary states 
are transformed into each other, and the third one is intact.\par
The bulk Ising CFT has $\mathbb{Z}_2$ symmetry which can be gauged, yielding the same CFT, that we call dual Ising CFT, -- historically known as the Kramers-Wannier duality. Correspondingly, following Ref.~\cite{hep-th/0201173}, we can implement $\mathbb{Z}_2$ gauging on the BCFT. Introduce a $\mathbb{Z}_2$ invariant boundary state by summing over a pair of the Cardy states which are transformed into each other under the $\mathbb{Z}_2$ symmetry:
\begin{equation}
\bket{f}_D=\frac{1}{\sqrt{2}}(\bket{+}_A+\bket{-}_A).\label{invariant d}
\end{equation}
 Also, we consider following boundary states
 \begin{eqnarray}
  \bket{+}_D&=&\frac{1}{\sqrt{2}}(\bket{f}_A+\kket{\eta}_A)\label{25}\\
\bket{-}_D&=&\frac{1}{\sqrt{2}}(\bket{f}_A-\kket{\eta}_A),\label{26}
 \end{eqnarray}
where, $\kket{\eta}_A$ is the Ishibashi state in the twisted sector, which is defined by $\kket{\eta}_A=\sqrt[4]{2}\kket{1/16}$. 
Two boundary states, $\bket{\pm}_D$ in Eqs.~(\ref{25}) and (\ref{26}) are called the fractional brane in the literature~\cite{hep-th/0201173}, see also Sec.~\ref{sec:dfixed} for more detailed discussion in a general case.
It is straightforward to show that three boundary states, $\bket{+}_D$, $\bket{-}_D$, and $\bket{f}_D$, satisfy the Cardy condition and by the form of the open string partition function, the boundary states $\bket{+}_D$, $\bket{-}_D$ behave as the fixed boundary states whereas $\bket{f}_D$ does as the free boundary state of the Ising CFT. 
Furthermore, after $\mathbb{Z}_2$ gauging of the boundary states, the fixed (resp. free) boundary state is mapped to free (resp. fixed) boundary state, $i.e.,$ the role of free and fixed boundary states is switched after gauging. This is in line with the fact that the original Ising CFT and $\mathbb{Z}_2$ gauged Ising CFT are dual with each other. \par  Fermionic Ising CFT is nothing but the free Majorana theory, realized in the critical Majorana chain~\cite{kitaev2001unpaired}. 
The boundary condition of the free Majorana theory is given by
  \begin{equation}
      (\psi_r-i\epsilon\overline{\psi}_{-r})\kket{B}=0,\;\;\epsilon=\pm1, r\in\mathbb{Z}+\frac{1}{2}\label{gluef}
  \end{equation}
 where $\psi_r$ (resp. $\overline{\psi}_{-r}$) is fermionic mode in the holomorphic (resp. anti-holomorphic) sector.  There are four states that satisfy (\ref{gluef}), and each two states is defined in the Neveu-Schwarz (NS) and the Ramond (R) sectors:
 \begin{eqnarray}
  \kket{\text{NS},\epsilon}&=&\displaystyle \prod_{r\in\mathbb{N}-1/2}e^{i\epsilon\psi_{-r}\overline{\psi}_{-r}}\Ket{0}_{\text{NS}}\label{NSboundary}\\
  \kket{\text{R},\epsilon}&=&\displaystyle \sqrt[4]{2} \prod_{r\in\mathbb{N}
  }e^{i\epsilon\psi_{-r}\overline{\psi}_{-r}}\Ket{\epsilon}_{\text{R}}\label{r}
 \end{eqnarray}
Here, $\Ket{0}_{\text{NS}}$ and $\Ket{\epsilon}_{\text{R}}$ is the ground state in the NS and R sector\footnote{The zero mode acts on the ground state on R sector as $\psi_0\Ket{\pm}_{\text{R}}=\frac{1}{\sqrt{2}}e^{\pm i\pi/4}\Ket{\mp}_{\text{R}}$, $\overline{\psi}_0\Ket{\pm}_{\text{R}}=\frac{1}{\sqrt{2}}e^{\mp i\pi/4}\Ket{\mp}_{\text{R}}$. }. These four states satisfy the Cardy condition as well as the condition of elementary boundary states except when two different boundary conditions are imposed on both ends in an open string partition function, namely, 
\begin{equation}
    S\Bigl[\bbra{\text{NS},+}\tilde{q}^{\frac{1}{2}(L_0+\overline{L}_0-1/24)}\kket{\text{NS},-}\Bigr]=\sqrt{2}\chi_{1/16}(q)
\end{equation}
Due to the presence of the prefactor $\sqrt{2}$, the four boundary states do not satisfy the Cardy condition. However, 
the prefactor $\sqrt{2}$ is an indication of ``a mod 2 anomaly''~\cite{2005.11314}, which is remedied by adding the Majorana zero mode by hand. 
Therefore, we generalise the notion of elementary boundary states to a fermionic case by allowing this prefactor $\sqrt{2}$ -- the four Ishibashi states in Eqs.~(\ref{NSboundary})(\ref{r}) are generalised elementary boundary states of the free fermion. \par

To relate the fermionic boundary states to the bosonic ones, we implement the GSO projection, mapping the fermionic boundary states to the ones which are invariant under $\mathbb{Z}_2$ fermion parity, $(-1)^{F+\overline{F}}$ ($F$ and $\overline{F}$ is the fermion number operator in the holomorphic and anti-holomorphic sector). The fermion parity $(-1)^{F+\overline{F}}$ acts trivially on the two Ishibashi states in the NS sector, whereas it does non-trivially on the ground states in the R sector in (\ref{r}); there are two ways to write the fermion parity operator of the zero mode in the ground state in the R sector,
\begin{equation}
    (-1)^{F+\overline{F}}=\begin{cases}
    -2i\psi_0\overline{\psi}_0\\
    2i\psi_0\overline{\psi}_0.\label{fer}
    \end{cases}
\end{equation}
In the first (resp. second) case, $\kket{\text{R},-}$ (resp. $\kket{\text{R},+}$ ) is invariant under the fermion parity. 
[Following Ref.~\cite{Bachas:2012bj}, we call the BCFT of the first and second choice of (\ref{fer}) $0$A theory and 0$B$ theory, which are named after type II$A$ and II$B$ string theory.]
Choosing the first case in (\ref{fer}) (the second case can be similarly discussed),
the GSO projection is executed as follows~\cite{Bachas:2012bj}. Define $(-1)^{F+\overline{F}}$ invariant boundary states with $\epsilon=-1$ as
\begin{equation}
    \kket{0A,\pm}=\frac{1}{\sqrt{2}}(\kket{\text{NS},-1}\pm \kket{\text{R},-1}),\label{0a2}
\end{equation}
as well as the one with $\epsilon=+1$
\begin{equation}
    \kket{0A}=\kket{\text{NS},+1}.\label{0a1}
\end{equation}
Notice that in the first case of (\ref{fer}), the boundary state $\kket{\text{R},+1}$
is projected out. By comparing open string partition functions of the GSO projected boundary states in Eqs.~(\ref{0a2}) and (\ref{0a1}) with those of the bosonic Ising CFT in (\ref{boson ising}), one can show that 
\begin{eqnarray}
 \kket{0A,+1}&=&\bket{+}_A\notag\\
 \kket{0A,-1}&=&\bket{-}_A\notag\\
 \kket{0A}&=&\bket{f}_A.
\end{eqnarray}
Therefore, 
we get
 \begin{equation}
     0A:\begin{cases}
     \kket{\text{NS},-1}=\frac{1}{\sqrt{2}}(\bket{+}_A+\bket{-}_A)\label{0a}\\
     \kket{\text{NS},+1}=\bket{f}_A\\
     \kket{\text{R},-1}=\frac{1}{\sqrt{2}}(\bket{+}_A-\bket{-}_A)
     \end{cases}
 \end{equation}
If we instead chose the second case of the fermion parity in (\ref{fer}), that is, if we adopt the 0$B$ theory, we obtain the similar relations to (\ref{0a}) with the sign in the fermionic boundary states inverted.
 
Eqs.~(\ref{0a}) is nothing but fermionization of boundary states of the Ising CFT. 
The simple example of the Ising BCFT has an important implication: $\mathbb{Z}_2$ orbifold of bosonic boundary states~[the first line of (\ref{0a})] and $\mathbb{Z}_2$ invariant boundary state, as shown in the second line of (\ref{0a}) constitute the fermionic boundary states in the NS sector, whereas $\mathbb{Z}_2$ odd bosonic boundary state, such as the third line of (\ref{0a}) yields the fermionic boundary state in the R sector. \par
In addition to fermionization of boundary states starting from the original bosonic Ising CFT ($A$-theory), we can also consider the fermionization of boundary states in the dual Ising theory ($D$-theory). Denoting fermionized theory from $A$-theory and $D$-theory by $F$-theory and $\tilde{F}$-theory, respectively, and choosing the first case for the fermion parity in (\ref{fer}), one finds
\begin{equation}
\begin{split}
    A\to F&:\begin{cases}
     \kket{\text{NS},-1}_F=\frac{1}{\sqrt{2}}(\bket{+}_A+\bket{-}_A)\\
     \kket{\text{NS},+1}_F=\bket{f}_A\\
     \kket{\text{R},-1}_F=\frac{1}{\sqrt{2}}(\bket{+}_A-\bket{-}_A)
     \end{cases}
     \\ D\to \tilde{F}&:\begin{cases}
     \kket{\text{NS},-1}_{\tilde{F}}=\frac{1}{\sqrt{2}}(\bket{+}_D+\bket{-}_D)\\
     \kket{\text{NS},+1}_{\tilde{F}}=\bket{f}_D\\
     \kket{\text{R},-1}_{\tilde{F}}=\frac{1}{\sqrt{2}}(\bket{+}_D-\bket{-}_D)
     \end{cases}\label{dd}
\end{split}
 \end{equation}
Referring to Eqs.~(\ref{invariant d})-(\ref{26}), (\ref{dd}) is refined to be
   \begin{equation}
   \begin{split}
        A\to F&:\begin{cases}
     \kket{\text{NS},-1}_F=\frac{1}{\sqrt{2}}(\bket{+}_A+\bket{-}_A)\\
     \kket{\text{NS},+1}_F=\bket{f}_A\\
     \kket{\text{R},-1}_F=\frac{1}{\sqrt{2}}(\bket{+}_A-\bket{-}_A)
     \end{cases}
     \\ A\to D\to \tilde{F}&:\begin{cases}
     \kket{\text{NS},-1}_{\tilde{F}}=\bket{f}_A\\
     \kket{\text{NS},+1}_{\tilde{F}}=\frac{1}{\sqrt{2}}(\bket{+}_A+\bket{-}_A)\\
     \kket{\text{R},-1}_{\tilde{F}}=\kket{\eta}_A
     \end{cases}
   \end{split}
  \end{equation}

In the following section, we see this procedure of fermionization of the boundary states is generically applicable to other CFTs with global $\mathbb{Z}_2$ symmetry.

\section{General case}\label{general}

We consider general CFTs with a global, non-anomalous $\mathbb{Z}_2$ symmetry.
A global non-anomalous $\mathbb{Z}_2$ symmetry allows one to either orbifold the theory or to fermionise it.
The precise object we are looking for here is the elementary boundary conditions/states. 
Since it will be important later, we reiterate what was reviewed in Sec. \ref{ising} to explain what they are.
Elementary boundary states \cite{hep-th/0108238} are defined so that the open-channel loop contains Virasoro characters with non-negative integer coefficients, and in particular that the coefficient of the vacuum character is unit for the same boundary state, and vanishing for the different, placed on two ends of the cylinder.
For fermionic theories, the presence of unpaired Majorana zero mode \cite{1912.01602,2005.11314,2006.07369} forces us to extend the definition of elementary a bit, so that the open-string character to be either $1$ or $\sqrt{2}$.
We will call it ``generalised elementary'', on par with what we defined in Sec. \ref{ising}.

\subsection{Orbifolding and fermionizing using \z2 defect lines}

We generalise the construction of \z2 orbifolding and fermionization on the torus to the construction of those on the cylinder.
In order to do this, we insert \z2 defect lines on the cylinder in all possible ways and combine them in a particular way \cite{1902.05550,2002.12283}.
We will hereafter employ the notation of \cite{2002.12283} and refer to the original theory as $A$-theory and the orbifolded theory as $D$-theory, while calling the two fermionic theories as $F$ and $\tilde{F}$.

\subsubsection{Orbifolding on the cylinder}

The untwisted (closed-string) sector of the $D$-type theory can be constructed from the $A$-type theory by summing up the \z2 defect configurations as follows,
\begin{align}
    \fig{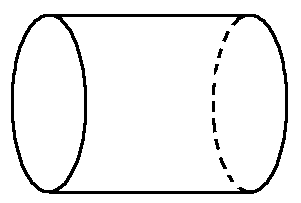}^{\rm untwisted}_{\text{$D$-type}}
    \equiv 
    \fig{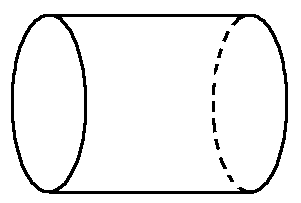}_{\text{$A$-type}}
    +\fig{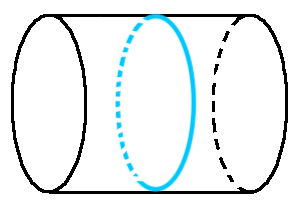}_{\text{$A$-type}}
    +\fig{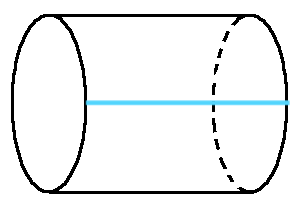}_{\text{$A$-type}}
    +\fig{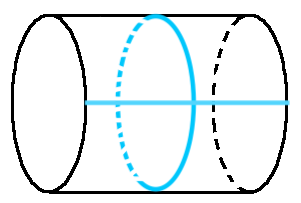}_{\text{$A$-type}}
    \label{eq:config}
\end{align}
while the twisted (closed-string) sector can be constructed by the sum,
\begin{align}
    \fig{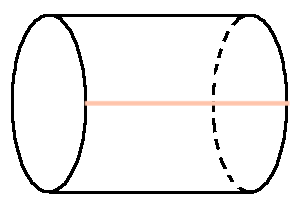}^{\rm twisted}_{\text{$D$-type}}
    \equiv 
    \fig{Ano}_{\text{$A$-type}}
    -\fig{Acentre}_{\text{$A$-type}}
    +\fig{Aline}_{\text{$A$-type}}
    -\fig{Acentreline}_{\text{$A$-type}}
\end{align}
Here, the blue lines, as in $\textfig{Aline}$, are the \z2 defect lines of the $A$-type theory, and the orange line, as in $\textfig{Dline}$, is the \z2 defect line of the $D$-type theory.
We will hereafter omit the subscript/superscript whenever it is apparent from context what specific theory or what specific sector of it we are talking about.

Note that the overall coefficient is different from the case of torus partition functions, which was $1/2$.
This is because the normalization of states are already fixed.\footnote{One can also say that this is because the genus of the cylinder is $0$. See \cite{1902.05550}.}
To see this concretely, consider the torus partition function of the orbifold theory and represent it in terms of that of the original theory,
\begin{equation}
\begin{split}
    &\Tr_{S^1}\left[e^{-\beta H_{\rm orbifold}}\right]\\
    =&\frac{1}{2}\left(\Tr_{S^1}\left[e^{-\beta H}\right]+\Tr_{S^1}\left[ge^{-\beta H_{\rm orbifold}}\right]+({\tt twisted})\right)
\end{split}
\end{equation}
where $g$ is our \z2 defect line and $({\tt twisted})$ refers to the contribution from the twisted sector of the original theory.
Concentrating on the first two and expanding the trace in terms of states, we have
\begin{equation}
 \begin{split}
    &\sum_{{n_{\rm orbifold}}}\Braket{n_{\rm orbifold}|e^{-\beta H_{{\rm orbifold}}}|n_{\rm orbifold}}\\
    =&\sum_{n\leq gn}\left[\left(\frac{\Bra{n}+\Bra{gn}}{\sqrt{2}}\right)e^{-\beta H}\left(\frac{\Ket{n}+\Ket{gn}}{\sqrt{2}}\right)
    +({\tt twisted})
    \right]
\end{split}
\end{equation}
where $n\leq gn$ means that we avoid summing over the same index twice.
It is therefore apparent that $\Ket{n_{\rm orbifold}}$ corresponds to $\frac{\Ket{n}+\Ket{gn}}{\sqrt{2}}$, or other similar looking states coming from the twisted sector of the original theory.
Let us now concentrate on the state coming from the untwisted sector of the original theory.
In terms of the closed-string amplitude and in terms of the above picture using the cylinder, this means we have
\begin{widetext}
\begin{equation}
\begin{split}
    &\Braket{n_{\rm orbifold}|\fig{Dno}\limits_{\text{$D$-type}}^{\rm untwisted}|m_{\rm orbifold}}
    \equiv
    \Braket{n_{\rm orbifold}|e^{-\beta H_{\rm orbifold}}|m_{\rm orbifold}}\\
    =&\left(\frac{\Bra{n}+\Bra{gn}}{\sqrt{2}}\right)e^{-\beta H}\left(\frac{\Ket{m}+\Ket{gm}}{\sqrt{2}}\right)
    =\Braket{n|e^{-\beta H}|m}+\Braket{n|e^{-\beta H}|gm}
    =\Braket{n|{\fig{Ano}}|m}+\Braket{n|{\fig{Acentre}}|m}\label{000}
    \end{split}
\end{equation}
\end{widetext}
where we used $g^2=1$ and the fact that $g$ commutes with the Hamiltonian and can be moved topologically on the cylinder.
If one wishes to generalise to states coming from the twisted sector of the original theory, one finally gets the picture we have given in the beginning.

\subsubsection{Fermionizing on the cylinder}

The NS sector of the $F$-type theory can be constructed from the $A$-type theory from the following sum,
\begin{align}
    \fig{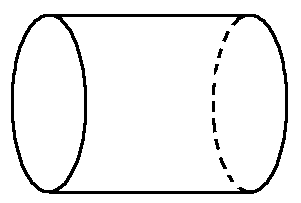}^{\rm NS}_{\text{$F$-type}}
    \equiv 
    \fig{Ano}_{\text{$A$-type}}
    +\fig{Acentre}_{\text{$A$-type}}
    +\fig{Aline}_{\text{$A$-type}}
    -\fig{Acentreline}_{\text{$A$-type}}
    \label{eq:FNS}
\end{align}
while the R sector can be constructed from the sum,
\begin{align}
    \fig{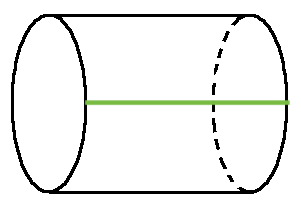}^{\rm R}_{\text{$F$-type}}
    \equiv 
    \fig{Ano}_{\text{$A$-type}}
    -\fig{Acentre}_{\text{$A$-type}}
    +\fig{Aline}_{\text{$A$-type}}
    +\fig{Acentreline}_{\text{$A$-type}}
    \label{eq:FR}
\end{align}
where the green line, as in $\textfig{Fline}$ represents the defect line flipping the fermion parity, $(-1)^{F}$.
The overall normalisation is determined using the same argument as in the orbifold case.

\subsection{Boundary states in the original theory}

In the untwisted sector of the original theory, one can classify boundary states into two groups, depending on whether they are \z2 symmetric or not.
We write the complete set of \z2 symmetric elementary boundary conditions as 
\begin{align}
    \mathcal{A}^{\rm free}\equiv \left\{\bket{f_{A,i}}\middle|i=1,2,\dots, N^{\rm free}_A\right\}
\end{align}
while that of \z2 breaking elementary boundary conditions as
\begin{align}
    \mathcal{A}^{\rm fixed}\equiv\mathcal{A}^{+}\bigcup \mathcal{A}^{-}\equiv\displaystyle\bigcup_{\pm}\left\{\bket{b_{A,i}^\pm}\middle|i=1,2,\dots, N^{\rm fixed}_A\right\}
\end{align}
where $\bket{b^+_i}$ maps to $\bket{b^-_i}$ under the \z2 symmetry and vice versa.
Here the subscript $A$ means we are considering the $A$-type theory.
Incidentally, although $N_A^{\rm free,fixed}$ are finite for rantional conformal field theories (RCFTs), those numbers lose their precise meaning for non RCFTs.
We speculate that the construction below works for non RCFTs as well at least when the number of conformal blocks is countably infinite.

On the other hand, in the twisted sector of the original theory, we denote the complete basis for boundary states (Ishibashi states in RCFTs) as
\begin{align}
    \tilde{\mathcal{A}}^{\rm twisted}\equiv \left\{\kket{\tau_i}\middle|i=1,2,\dots, N^{\rm twisted}_A\right\},
\end{align}
or, if we are interested in the complete set of elementary boundary states, we can write it as 
\begin{align}
    {\mathcal{A}}^{\rm twisted}\equiv \left\{\bket{\tau_i}\middle|i=1,2,\dots, N^{\rm twisted}_A\right\},
\end{align}
where $N_A^f\equiv N^{\rm free}_{A,D}= N^{\rm twisted}_{A,D}$ as we will explain in Appendix \ref{sec:same}.
Some comments are in order.
\begin{itemize}
    \item Unlike the untwisted case, the open string spectrum of any linear combination of those states will always contain Virasoro characters with negative integer from the twisted sector. 
    The notion of ``elementary'' here is therefore slightly generalised to accommodate for such negative integers.
    \item More precisely, we will define $\mathcal{A}^{\rm twisted}$ using the information that the $D$-type symmetry breaking boundary states are elementary.
    See later subsections for more explanation.
    \item The boundary states in $\mathcal{A}^{\rm twisted}$ are all invariant under the \z2 transformation.
    The reason is the following: if we have such a twisted boundary state $\bket{\mathcal{O}}$ that it is not invariant under \z2, then we have an operator $(\mathcal{O},\overline{\mathcal{O}})$ in the \z2 odd twisted sector.
    Meanwhile, under fermionization, the \z2 odd twisted sector maps to the fermionic NS sector --
    This means that the operator $(\mathcal{O},\overline{\mathcal{O}})$ must have a half-integer spin, which is obviously a contradiction.\footnote{
    We can also see this mathematically from the fact that the twisted odd torus partition function is given by $Z_{\rm twisted}-\mathcal{T}Z_{\rm twisted}$, and that $\abs{\chi_i}$ is invariant under the action of $\mathcal{T}$. 
    }
\end{itemize}

\subsection{Boundary states for the orbifold theory}

Let us now consider the elementary boundary states in the $D$-theory.
Although the result has already been reproduced many times before \cite{hep-th/0201173,hep-th/0108126,hep-th/9708141,hep-th/0011060}, it is instructive to compare with the fermionic case.

The general idea here is the following.
First consider cylinder amplitudes between elementary boundary states of the $A$-theory.
By virtue of \it e.g., \rm \eqref{eq:config}, we get the cylinder amplitude of the $D$-theory, shown on the LHS.
We can then use this cylinder amplitude to infer the boundary states which reproduce it.

\subsubsection{\z2 breaking boundary conditions}
\label{sec:Dbreaking}

Let us first study the untwisted closed string amplitude in the $D$-theory.
Sandwiching the RHS of \eqref{eq:config} by $\bket{b_{A,i}^{+}}$, we get\footnote{The argument will be the same when we have two different boundary conditions on two ends.}
\begin{equation}
\begin{split}
    &\Braket{\widehat{b_{A,i}^{+}}|\fig{Ano}+\fig{Acentre}+\fig{Aline}+\fig{Acentreline}|\widehat{b_{A,i}^{+}}}\\
    =
    &\Braket{\widehat{b_{A,i}^{+}}|q^{\frac{1}{2}\left(L_0+\bar{L}_0-\frac{c}{12}\right)}|\widehat{b_{A,i}^{+}}}+\Braket{\widehat{b_{A,i}^{+}}|q^{\frac{1}{2}\left(L_0+\bar{L}_0-\frac{c}{12}\right)}|\widehat{b_{A,i}^{-}}}\label{sandwich},
\end{split}
\end{equation}
where we used that the last two terms are vanishing, since $\bket{b_{A,i}^{\pm}}$ is in the untwisted sector of the original theory.
We then look for boundary states which correctly reproduces this cylinder partition function, which turn out to be
\begin{align}
    \frac{1}{\sqrt{2}}\left(\bket{b_{A,i}^{+}}+\bket{b_{A,i}^{-}}\right),
\end{align}
so that they are the boundary states (which might or might not be elementary unless we check) of the $D$-theory.

Likewise, for the twisted sector of the $D$-theory, the cylinder partition function
\begin{align}
    \Braket{\widehat{b_{A,i}^{+}}|q^{\frac{1}{2}\left(L_0+\bar{L}_0-\frac{c}{12}\right)}|\widehat{b_{A,i}^{+}}}-\Braket{\widehat{b_{A,i}^{+}}|q^{\frac{1}{2}\left(L_0+\bar{L}_0-\frac{c}{12}\right)}|\widehat{b_{A,i}^{-}}},
\end{align}
is reproduced by a boundary state
\begin{align}
    \frac{1}{\sqrt{2}}\left(\bket{b_{A,i}^{+}}-\bket{b_{A,i}^{-}}\right),
\end{align}
and they are also the boundary states (which might or might not be elementary unless we check) of the $D$-theory, this time in the twisted sector.

At this stage, these might not be elementary boundary conditions, and can be a sum of other 
elementary boundary conditions with integer coefficients.
Here, to reiterate, the elementary boundary states must reproduce the Virasoro character inside the open-string spectrum with integer coefficients, non-negative for the untwisted, and possibly negative for the twisted sector of the theory.
One can indeed check that this is so for states above among themselves, so that we conclude that these are elementary boundary conditions for the $D$-theory.


\subsubsection{\z2 symmetric boundary conditions}\label{sec:dfixed}

By using the same argument as above, we can see that the closed string amplitude in the untwisted sector can be reproduced by the state
\begin{align}
    \frac{1}{\sqrt{2}}\left(\bket{f_{A,i}}+\bket{f_{A,i}}\right)=\sqrt{2}{\bket{f_{A,i}}},
    \label{eq:sumfree}
\end{align}
or
\begin{align}
    \frac{1}{\sqrt{2}}\left(\bket{\tau_{A,i}}+\bket{\tau_{A,i}}\right)=\sqrt{2}{\bket{\tau_{A,i}}},
    \label{eq:sumfreetwist}
\end{align}
and that there are no corresponding boundary states in the twisted sector.
Note that $\sqrt{2}{\bket{f_{A,i}}}$ is invariant under the dual \z2, while $\sqrt{2}{\bket{\tau_{A,i}}}$ flips the sign under it, since they came from the twisted sector of the $A$-theory.

The question is if they are really elementary boundary states.
This time, one can easily see that the states are not elementary since they produces two times the vacuum character in the open-string channel.
In other words, these states like $\sqrt{2}{\bket{f_{A,i}}}$ or $\sqrt{2}{\bket{\tau_{A,i}}}$ should be a sum/subtraction of two elementary boundary states.
By demanding that they are elementary, as well as that they reproduce the boundary state of the original $A$-theory when dualised again, we can see that the only consistent choice 
is
\begin{align}
    \frac{1}{\sqrt{2}}\left(\bket{f_{A,i}}\pm\bket{\tau_{A,i}}\right).
    \label{eq:Dfixed}
\end{align}
The other choices like $\frac{1}{\sqrt{2}}\left(\bket{f_{A,i}}\pm\bket{f_{A,i}}\right)$ or $\frac{1}{\sqrt{2}}\left(\bket{\tau_{A,i}}\pm\bket{\tau_{A,i}}\right)$ are eliminated, because free boundary conditions must correspond to fixed boundary conditions under duality.\footnote{We thank Yunqin Zheng for private communication of the result.}
Note that when the system has an enhanced symmetry, it might also be possible to rotate boundary states with the same Virasoro dimension to the above forms --
The result merely means that there must exist a canonical form \eqref{eq:Dfixed}, for the $D$-theory elementary boundary states.
Incidentally, they are called the fractional branes in the context of string theory \cite{hep-th/0201173}.
It is also consistent with $N_{A}^{\rm fixed}=N_{A}^{\rm twisted}$.

Although we have already prepared the twisted sector boundary states $\bket{\tau_{A,i}}$ in the $A$-theory, one should understand this as a way of determining $\bket{\tau_{A,i}}$ in terms of $\kket{\tau_{A,i}}$, by requiring that \eqref{eq:Dfixed} be elementary in the $D$-theory.
More concretely, one can write
\begin{align}
    \bket{\tau_{A,i}}= \sum_{i}\frac{C_{ij}}{\sqrt{S_{0}^{{i}}}} \kket{\tau_{A,j}} \quad \text{where} \quad \sum_{j}\abs{C_{ij}}^2=1,
\end{align}
and we determine $C_{ij}$, demanding that \eqref{eq:Dfixed} are elementary boundary states.
For the simplest case where there is only one such $\kket{\tau}$, the coefficient is simply $C=1$.
The coefficients for more general case has been determined using classifying algebra in \cite{hep-th/9708141,hep-th/9712257, hep-th/9902132,hep-th/9908025}.
In this context, the matrix $C_{ij}$ is related to the so-called fixed-point resolution matrix.

\subsubsection{Final result and its consistency}

We conclude that for the $D$-type theory, the complete set of elementary boundary conditions becomes the following:
\begin{align}
\begin{split}
    \mathcal{D}^{\rm free}&= \left\{\frac{1}{\sqrt{2}}\left(\bket{b_{A,i}^{+}}+\bket{b_{A,i}^{-}}\right)\middle|i=1,2,\dots, N^{\rm fixed}_A\right\}\\
        \mathcal{D}^{\rm fixed}&\equiv\mathcal{D}^{+}\bigcup \mathcal{D}^{-}\\&=\bigcup_{\pm}\left\{\frac{1}{\sqrt{2}}\left(\bket{f_{A,i}}\pm\bket{\tau_{A,i}}\right)\middle|i=1,2,\dots, N^{f}_A\right\}
    \\
    {\mathcal{D}}^{\rm twisted}&= \left\{\frac{1}{\sqrt{2}}\left(\bket{b_{A,i}^{+}}-\bket{b_{A,i}^{-}}\right)\middle|i=1,2,\dots, N^{\rm fixed}_A\right\}
\end{split}
\end{align}
If the boundary states are fixed or free can be easily determined, by noting that the sign of $\bket{\tau_{A,i}}$ is flipped under the dual \z2. 
This is also consistent with the general analysis that fixed boundary conditions map to free ones and vice versa under orbifolding.\footnote{One instance of this is the mapping of Neumann and Dirichlet boundary conditions under the particle-vortex duality.}


As a first consistency check, one can see that the total number of elementary boundary states are unchanged by orbifolding, as they should be.
Also, as we explained previously, we trivially get integer coefficient sums of open-channel Virasoro characters for boundary conditions among boundary states in $\mathcal{D}^{\rm free}$ or $\mathcal{D}^{\rm twisted}$.
Incidentally, the states in $\mathcal{D}^{\rm fixed}$
are already elementary since we prepared them to be so, by tuning $C_{ij}$ --
The fact that we have coefficients to tune was crucial in finding the elementary boundary states of the $D$-theory.

\subsection{Boundary states for the fermionic theory}

In the most part of this section, we will choose a specific convention for the parity symmetry and look at the ${F}$-type theory as defined in \cite{1905.08943}.
We will also briefly discuss the $\tilde{F}$-type theory for consistency.

\subsubsection{NS sector boundary conditions}

For fermionic theories, we can just mimic the steps we discussed in the orbifold theory.
In the NS sector, the cylinder partition function \eqref{eq:FNS} can be reproduced by boundary states from $\bket{b^\pm_{A,i}}$ and from $\bket{f_{A,i}}$, which are both even under $(-1)^F$,
\begin{align}
    \begin{cases}
        \displaystyle\frac{1}{\sqrt{2}}\left(\bket{b^+_{A,i}}+\bket{b^-_{A,i}}\right) & \text{even under $(-1)^F$}\\
        \sqrt{2}\bket{f_{A,i}} & \text{even under $(-1)^F$}
    \end{cases}
\end{align}
Up to this point, the argument is almost similar to that of the $D$-theory, except that there are no states from $\bket{\tau_{A,i}}$ in the NS sector, while there was for the $D$-theory.
Also, the convention of the parity symmetry is so that the first states are odd under $(-1)^F$, while the second being even.

As we discussed in the previous subsection, the state $\sqrt{2}\bket{f_{A,i}}$ is not an elementary boundary condition in its strict sense of the word.
For example, if we place the former and the latter boundary states on two ends of a cylinder, the open string spectrum we read out comes with $\sqrt{2}$ times integer coefficients.
As discussed in {\cite{1912.01602,2005.11314,2006.07369}}, this is an indication that we need to add a boundary Majorana fermion by hand, to cancel the anomaly created by the Majorana zero mode, or equivalently, that these two boundary states are incompatible.

\subsubsection{R sector boundary states}

In the R sector, 
the cylinder partition function \eqref{eq:FR} can be reproduced by boundary states coming from $\bket{b^\pm_{A,i}}$ and from $\bket{\tau_{A,i}}$,
\begin{align}
    \begin{cases}
        \displaystyle\frac{1}{\sqrt{2}}\left(\bket{b^+_{A,i}}-\bket{b^-_{A,i}}\right) & \text{odd under $(-1)^F$}\\
        \sqrt{2}\bket{\tau_{A,i}} & \text{even under $(-1)^F$}
    \end{cases}
    \label{eq:lo}
\end{align}
The $\sqrt{2}$ here requires even more attention than the NS case.
At first we would imagine that this comes from the Majorana zero mode again, but this cannot be the case here -- The zero mode will multiply the $R$-sector partition function by zero.
The other possibility is that they appear as a combination of two elementary boundary states, but this does not happen generically.
We can therefore only conclude that states like $\sqrt{2}\bket{\tau_{A,i}}$ does not appear in the fermionic theory.
This does not sound very symmetric, but it is not a contradiction. As we will explain below, the boundary states $\bket{\tau_{A,i}}$ instead become elementary for the $\tilde{F}$-type theory.

\subsubsection{Final result and its consistency}

We conclude that for the ${F}$-type theory, the complete set of elementary boundary conditions becomes
\begin{align}
    \mathcal{{F}}^{\rm NS}_{1}&= \left\{\frac{1}{\sqrt{2}}\left(\bket{b_{A,i}^{+}}+\bket{b_{A,i}^{-}}\right)\middle|i=1,2,\dots, N^{\rm fixed}_A\right\}\\
    \mathcal{{F}}^{\rm NS}_{\sqrt{2}}&= \left\{{\sqrt{2}}\bket{f_{A,i}}\middle|i=1,2,\dots, N^{f}_A\right\}\\
    \mathcal{{F}}^{\rm R}&= \left\{\frac{1}{\sqrt{2}}\left(\bket{b_{A,i}^{+}}-\bket{b_{A,i}^{-}}\right)\middle|i=1,2,\dots, N^{\rm fixed}_A\right\}
\end{align}
where two different groups of NS sector elementary boundary conditions were shown to be incompatible with each other.
The subscripts, $1$ and $\sqrt{2}$, indicate that the ground state degeneracy is $1$ or $2$ respectively, when we put the same boundary condition on the two ends.
The result is consistent with the one in \cite{2001.05055}, where several examples are considered. 

For consistency and completeness, let us also study the boundary conditions for the $\tilde{F}$-type theory.
In order to do this, one can simply apply the same procedure to the $D$-theory, where we get
\begin{align}
    \mathcal{\tilde{F}}^{\rm NS}_{\sqrt{2}}&= \left\{\bket{b_{A,i}^{+}}+\bket{b_{A,i}^{-}}\middle|i=1,2,\dots, N^{\rm fixed}_A\right\}\\
    \mathcal{\tilde{F}}^{\rm NS}_{1}&= \left\{\bket{f_{A,i}}\middle|i=1,2,\dots, N^{f}_A\right\}\\
    \mathcal{\tilde{F}}^{\rm R}&= \left\{\bket{\tau_{A,i}}\middle|i=1,2,\dots, N^{f}_A\right\}
\end{align}
One can see that the boundary states which we threw away in the $F$-type theory is recovered in the $\tilde{F}$-type.
This is not surprising, since $F$- and $\tilde{F}$-type theories are in general different, even though the difference is just a result of the discrete choice in $(-1)^F$.

\section{Discussions and future directions}\label{sec:discussion}

We have studied the complete set of elementary boundary states for two-dimensional fermionic CFTs.
In deriving them, we first used the construction of orbifold or fermionic theories using \z2 defect lines, and computed the closed string amplitude for the target theories, and then inferred the consistent boundary states for them.
We took maximal precaution that the resulting boundary states actually reproduce the open-string spectrum after $S$-transformation.
The consistency for the orbifold theory is that the coefficients in front of the open-channel Virasoro characters are integers, while for the fermionic theory we also allowed for $\sqrt{2}$ times the integer.
We also argue that there are two classes of boundary states in the fermionic theory, because of the $\sqrt{2}$ above, mutually inconsistent with each other when placed on two ends of a cylinder.
We have in the end found a consistent reshuffling of boundary states, as one moves from the $A$-type to $D$-, $F$-, or $\tilde{F}$-type theories.
The final result is summarised in Table \ref{table:only}.

\begin{table}[htb]
\caption{The final result on the complete set of the elementary boundary states in two-dimensional CFTs. Free/fixed indicates whether the boundary conditions are free/fixed. $\mathbb{Z}$ or $\sqrt{2}\mathbb{Z}$ indicates in which of the two groups of mutually incompatible boundary conditions they are.}
\begin{subtable}{\linewidth}
\centering
{
\begin{tabular}[t]{>{\centering}p{10mm}|>{\centering}p{47.5mm}|>{\centering\arraybackslash}p{24mm}}
  $A$-type & untwisted & twisted \\\hline\hline
  free & $\bket{f_a}$ & $\bket{\tau_a}$\\\hline
  fixed & $\bket{b_i^+}\xleftrightarrow[\mathbb{Z}_2]{} \bket{b_i^+}$ & N/A
\end{tabular}
}
\end{subtable}

\vspace{0.1cm}

\begin{subtable}{\linewidth}
\centering
{
\begin{tabular}[t]{>{\centering}p{10mm}|>{\centering}p{47.5mm}|>{\centering\arraybackslash}p{24mm}}
 $D$-type &   untwisted & twisted \\\hline\hline
  free & $\dfrac{1}{\sqrt{2}}\left(\bket{b_i^+}+\bket{b_i^-}\right)$ & $\dfrac{1}{\sqrt{2}}\left(\bket{b_i^+}-\bket{b_i^-}\right)$ \\\hline
  fixed & $\dfrac{1}{\sqrt{2}}\left(\bket{f_a}+\bket{\tau_a}\right)\xleftrightarrow[\mathbb{Z}_2]{}\dfrac{1}{2}\left(\bket{f_a}-\bket{\tau_a}\right)$ & N/A
\end{tabular}
}
\end{subtable}

\vspace{0.1cm}

\begin{subtable}{\linewidth}
\centering
{
\begin{tabular}[t]{>{\centering}p{10mm}|>{\centering}p{47.5mm}|>{\centering\arraybackslash}p{24mm}}
 $F$-type &   NS & R \\\hline\hline
  $\mathbb{Z}$ & $\dfrac{1}{\sqrt{2}}\left(\bket{b_i^+}+\bket{b_i^-}\right)$ & $\dfrac{1}{\sqrt{2}}\left(\bket{b_i^+}-\bket{b_i^-}\right)$\\\hline
  $\sqrt{2}\mathbb{Z}$ & $\sqrt{2}\bket{f_{a}}$ & N/A
\end{tabular}
}
\end{subtable}

\vspace{0.1cm}

\begin{subtable}{\linewidth}
\centering
{
\begin{tabular}[t]{>{\centering}p{10mm}|>{\centering}p{47.5mm}|>{\centering\arraybackslash}p{24mm}}
 $\tilde{F}$-type &   NS & R \\\hline\hline
  $\mathbb{Z}$ & $\bket{f_{a}}$ & $\bket{\tau_a}$\\\hline
  $\sqrt{2}\mathbb{Z}$ & $\bket{b_i^+}+\bket{b_i^-}$ & N/A
\end{tabular}
}
\end{subtable}
\label{table:only}
\end{table}

There are several interesting future directions to consider.
First of all, the detailed analysis of the R sector boundary conditions are necessary.
It would be interesting to find examples where latter states in \eqref{eq:lo} is actually constructed as a sum of two elementary boundary states like $\frac{1}{\sqrt{2}}\left(\bket{\tau_{A,i}}\pm \bket{\tau_{A,j}}\right)$.

Second, it is worthwhile to study other symmetries preserved by boundary states of fermionic theories.
It had been a folklore \cite{1704.01193} and has been proven recently \cite{2012.15861} that the existence of a symmetry preserving boundary condition implies the vanishing of the anomaly, for very general classes of symmetries and systems.
One can therefore now reverse the logic and use boundary states as a probe of anomaly in two-dimensional CFTs, or to study three-dimensional bosonic/fermionic SPT phases.
For example, it would be interesting to extend the analysis of \cite{2102.02203} to find fermionic theories with seperate holomorphic and anti-holomorphic parity symmetries, and study their anomalies.
It would also be interesting to see if the reverse statement is always true, as proposed in \cite{1904.06924,1908.02918} (See also \cite{1712.09361}).


Realization of the fermionic boundary states in a UV lattice model is also an important issue. If we introduce a finite (bosonic) spin chain with with an appropriate boundary condition by imposing a local magnetic field on a boundary, then we would obtain an eigenstate corresponding to a (bosonic) conformal boundary state. Using these eigenstates, together with our results would give rise to the desired Fermioninc boundary states. This strategy could work in e.g., a spin-$k/2$ chain whose criticality is governed by $SU(2)_k$ WZW CFT and the Potts models
(see \cite{2002.12283} where the fermionization of bulk of the three states Potts model was numerically confirmed.).
One could also construct the fermionic boundary state in a 2D classical statistical lattice model where defect lines is explicitly described. As an example, in the classical 2D Ising lattice model, the work \cite{aasen2016topological} introduced the $\mathbb{Z}_2$ defect lines
which is crucial in our argument as seen from the diagram in e.g,~\eqref{000}. 
\par

It will also be interesting to do the same for supersymmetric theories in two-dimensions, so that one knows the complete set of supersymmetric boundary conditions and/or defects.
These pieces of information are to be used in supersymmetric localisation computation on the hemisphere or with defects.
This direction has already been started in \cite{0809.0175} and continued in \cite{1703.09148,2001.05055}.

Higher-dimensional generalisation is also interesting. Although one cannot easily classify boundary conditions in higher-dimensions, it is already known that the particle-vortex duality maps Dirichlet type boundary conditions to Neumann type boundary conditions in some examples \cite{1712.02801,1803.08507,1902.09567,2012.07733}.
It is interesting to find other examples of equivalent boundary conditions across duality.
It is also intriguing to understand if there is an example of mutually incompatible boundary conditions in higher dimensions.

Lastly, it might also be interesting to consider cases where the duality itself is anomalous, for example the $S$-duality of electromagnetic theories \cite{hep-th/9505186,1905.08943}.
Anomaly of duality can also affect physical quantities as the anomaly of ordinary symmetries do.
For example, the partition function of free Maxwell theory in four dimensions are known to transform as a modular form under $S$-duality, rather than being invariant because of the duality anomaly \cite{hep-th/9505186}.
Another example, which pertains further to the presence of boundaries is the entanglement entropy -- 
the R\'enyi entropy of Abelian $p$-form free electromagnetic theory is known not to be duality invariant, and this phenomenon was called the entanglement anomaly and afflicted to the duality anomaly in \cite{1611.05920}.
The authors used electric boundary conditions \cite{1312.1183} on both sides of the duality to compute entanglement entropy to argue this --
In light of the present paper, what one should be doing first is to look for the boundary condition for the dual theory, which is dual to the electric boundary condition of the original theory, which was done for $\mathcal{N}=4$ SYM in \cite{0807.3720} and for QED in four dimensions in \cite{1902.09567}.

\medskip

\noindent \textbf{Note added:} Before the completion of this paper, \cite{2102.02203} appeared which contains results we present in this paper, deriving boundary conditions of all fermionic minimal models.
The paper also lists all the fermionic minimal models with holomorphic parity symmetry, but this is beyond the scope of the present paper.
We also like to mention another paper \cite{Fukusumi:2021zme} by Yoshiki Fukusumi, Yuji Tachikawa, and Yunqin Zheng, with whom we discussed together in the early stages of both work.
We also coordinated the publication of the two.

\section*{Acknowledgements}
The authors are grateful to Yoshiki Fukusumi, Rohit R. Kalloor, Ryohei Kobayashi, Yuji Tachikawa, and Yunqin Zheng for discussions.
We also thank Yoshiki Fukusumi, Yuji Tachikawa, and Yunqin Zheng for collaborations at the early stages of the work, and for the coordination of the publication of \cite{Fukusumi:2021zme} with the present paper.
We also especially thank Rohit for detailed discussions and collaborations throughout the entire stage of the project.
The work of HE is supported by the Koshland postdoc fellowship.
The work of MW is supported by
the Foreign Postdoctoral Fellowship Program of the Israel Academy of Sciences and Humanities.
MW also thanks Yukawa Institute for Theoretical Physics for support and hospitality while this work was in progress.
This work was in part supported by an Israel Science Foundation center for excellence grant (grant number 1989/14), by the Minerva foundation with funding from the Federal German Ministry for Education and Research, the ERC under the European Union’s Horizon 2020 research and innovation programme (grant agreement LEGOTOP No 788715), and the CRC/Transregio 183.

\appendix

\section{The proof that $N^{\rm free}=N^{\rm twisted}$}
\label{sec:same}

We prove that the number of free boundary states in the untwisted sector and that of twisted sector boundary states are the same.\footnote{We thank Ryohei Kobayashi for discussions.}
The action of the Verlinde line $\mathcal{L}_k$ to the CFT state can be written as
\begin{align}
    \mathcal{L}_k\Ket{\phi_\ell}=\frac{S_{k\ell}}{S_{0\ell}}\Ket{\phi_\ell}
\end{align}
In particular, when a system has a \z2 symmetry, there exist $\mathcal{L}_{\mathbb{Z}}$ such that
\begin{align}
    \mathcal{L}_{\mathbb{Z}}\Ket{\phi_\ell}=
    \begin{cases}
        \Ket{\phi_\ell} & \text{when $\phi_\ell$ is even}\\
        -\Ket{\phi_\ell} & \text{when $\phi_\ell$ is odd}
    \end{cases}
\end{align}
In other words, 
\begin{align}
    \frac{S_{\mathbb{Z}_2,\ell}}{S_{0,\ell}}=
    \begin{cases}
        1 & \text{when $\phi_\ell$ is even}\\
        -1 & \text{when $\phi_\ell$ is odd}
        \label{eq:imp}
    \end{cases}
\end{align}

Meanwhile, the twisted sector partition function can be written as
\begin{align}
    Z_{\rm twisted}=\sum_{i,j} N_{\mathbb{Z}_2,i}^{j} \chi_i \bar\chi_j
\end{align}
where 
\begin{align}
    N_{\mathbb{Z}_2,i}^{j}\equiv \sum_{\ell}\frac{S_{\mathbb{Z}_2,\ell}S_{i,\ell}\bar{S}_{i,\ell}}{S_{0,\ell}}
\end{align}

Now, the free boundary condition $\bket{f}$ is defined such that the open-string spectrum (the $S$-transformation of $\chi_f$) only contains the \z2 even primaries.
This means that
\begin{align}
    S_{f,\ell}
    =
    \begin{cases}
        1 & \text{when $\phi_\ell$ is even}\\
        0 & \text{when $\phi_\ell$ is odd}
    \end{cases}
\end{align}

Let us first prove that the there is a corresponding boundary state in the twisted sector for every free boundary state in the untwisted sector, \it i.e., \rm $N_{\mathbb{Z},f_j}^{f_k}=\delta_{i,j}$.
This can be proven as
\begin{equation}
\begin{split}
    N_{\mathbb{Z},f_j}^{f_k}
    &=
    \sum_{\ell}\frac{S_{\mathbb{Z}_2,\ell}S_{f_j,\ell}\bar{S}_{f_k,\ell}}{S_{0,\ell}}\\
    &=\sum_{\ell\in \text{even}}\frac{S_{\mathbb{Z}_2,\ell}S_{f_j,\ell}\bar{S}_{f_k,\ell}}{S_{0,\ell}}
    =\sum_{\ell\in \text{even}}{S_{f_j,\ell}\bar{S}_{f_k,\ell}}=\delta_{i,j}
\end{split}
\end{equation}
The inverse is also true. What we want to prove is that when $N_{\mathbb{Z}_2, i}^{i}=1$, we have $S_{i,\ell}=0$ for odd $\phi_\ell$.
We have the relation
\begin{align}
    1=N_{\mathbb{Z},i}^{i}
    =\sum_{\ell\in \text{even}}{S_{i,\ell}\bar{S}_{i,\ell}}
    -\sum_{\ell\in \text{odd}}{S_{i,\ell}\bar{S}_{i,\ell}},
\end{align}
but since we also have from unitarity that
\begin{align}
    1=
    \sum_{\ell\in \text{even}}{S_{i,\ell}\bar{S}_{i,\ell}}
    +\sum_{\ell\in \text{odd}}{S_{i,\ell}\bar{S}_{i,\ell}}
\end{align}
we have proven that $S_{i,\ell}=0$ for odd $\phi_\ell$.

\section{Fixing the fixed boundary states in the $D$-theory}
\label{sec:why}

{In this appendix, we show why \eqref{eq:Dfixed} is the correct form of the boundary state in $D$-theory by eliminating other possible construction of elementary boundary states. }
Let us split \eqref{eq:sumfree}, pretending that we do not know the final answer for a moment as $\sqrt{2}{\bket{f_{A,i}}}=\bket{f_{A,i}^+}+\bket{f_{A,i}^-}$
or in other words,
\begin{align}
    \bket{f_{A,i}^\pm}\equiv \frac{1}{\sqrt{2}}\left(\bket{f_{A,i}}\pm\bket{\pey_i}\right)
\end{align}
Let us now assume that they are invariant under the dual \z2 symmetry and see that it leads to contradiction.
If these are invariant under the dual \z2, by again orbifolding the theory, we will have untwisted sector boundary states, which are again sum/subtraction of elementary boundary states. 
They look like
\begin{align}
    \bket{f_{A,i}}\pm\bket{\pey_i},
\end{align}
and so $\bket{\pey_i}$ must also be an elementary boundary state in the untwisted sector.
The only possibility where it can happen is that $\bket{\pey_i}$ is actually one of $\left\{\bket{f_{A,i}}\right\}$.

Now we have seen that the candidate for the elementary boundary states is
\begin{align}
    \frac{1}{\sqrt{2}}\left(\bket{f_{A,i}}\pm\bket{f_{A,j}}\right),
\end{align}
and by similar arguments, and because the numbers of $\bket{f_{A,i}}$ and of $\bket{\tau_{A,i}}$ are the same, we also have boundary states like
\begin{align}
    \frac{1}{\sqrt{2}}\left(\bket{\tau_{A,i}}\pm\bket{\tau_{A,j}}\right).
\end{align}
However, this means that by orbifolding we get untwisted sector boundary states,
\begin{align}
    \bket{\tau_{A,i}}\pm\bket{\tau_{A,j}}.
\end{align}
Since $\bket{\tau_{A,i}}$ are in the twisted sector, this is a contradiction.

We have therefore proven that \eqref{eq:Dfixed} is the correct choice for the $D$-type boundary states, modulo a slight caveat that when the theory has a larger symmetry than \z2, free rotations of boundary states with the same Virasoro dimension can connect \eqref{eq:Dfixed} to the above ``pathological'' boundary states.

\bibliographystyle{apsrev4-1}
\bibliography{ref,main}

\end{document}